\documentclass[mathleft
]{an}
\usepackage{graphicx}
\usepackage{times}
\begin{document}

\Pagespan{789}{}
\Yearpublication{2008}%
\Yearsubmission{2008}%
\Month{11}%
\Volume{999}%
\Issue{88}%

\title{Galaxy mergers at high resolution:\\ 
From elliptical galaxies to tidal dwarfs and globular clusters}

\author{F. Bournaud\inst{1}\fnmsep\thanks{Corresponding author:
  \email{frederic.bournaud@cea.fr}\newline}, M. Bois\inst{2}, E. Emsellem\inst{2}, P.-A. Duc\inst{3}
}
\titlerunning{Galaxy mergers at high resolution}
\authorrunning{F. Bournaud et al.}
\institute{
CEA, IRFU, SAP. F-91191 Gif-sur-Yvette, France. 
\and
   Universit\'e de Lyon, Universit\'e Lyon~1, Lyon, 
   Observatoire de Lyon, 9 av. Charles Andr\'e, Saint-Genis
   Laval, F-69230; CNRS, UMR 5574, Centre de Recherche Astrophysique
   de Lyon ; Ecole Normale Sup\'erieure de Lyon, France
\and
Laboratoire AIM, CEA/DSM -- CNRS -- Universit\'e Paris Diderot. CEA Saclay, IRFU/SAp, F-91191 Gif-sur-Yvette, France. 
}

\received{xx xxx 2008}
\accepted{xx xxx 2008}
\publonline{later}

\keywords{galaxies: interactions -- galaxies: evolution -- galaxies: elliptical -- globular clusters: general}

\abstract{ Numerical simulations of galaxy mergers are a powerful tool to study these fundamental events in the hierarchical built-up of galaxies. Recent progress have been made owing to improved modeling, increased resolution and large statistical samples. We present here the highest-resolution models of mergers performed so far. The formation of a variety of substructures ranging from kinematically decoupled cores to globular-like clusters is directly resolved. In a resolution study, we show that the large-scale structure of elliptical-like merger remnants can be affected by the resolution, and a too modest resolution may affect the numerical predictions on the properties of major merger remnants: understanding precisely which kind of event or succession of events has formed the various types of elliptical galaxies remains an open challenge. }

\maketitle
\sloppy

\section{Introduction}
For more than two decades, numerical simulations have been a powerful tool to study the dynamics of interacting galaxies and the properties of galaxy merger remnants. Some of the main results obtained from numerical models range from the formation of elliptical-like galaxies with realistic $r^{1/4}$ density profiles in major mergers of massive spirals (Barnes 1992, Naab \& Burkert 2003, Bournaud et al. 2005), the effect of minor mergers of spiral galaxies with small companions (e.g., Walker et al. 1996), and the triggering of star-formation by galaxy interactions and mergers (Mihos \& Hernquist 1994, Cox et al. 2006, di~Matteo et al. 2007).

Recent progress have been made by studying galaxy interactions in a wider range of physical conditions including more complex (and more realistic) cases like:  the (re-)merging of early-type galaxies that are themselves remnants of past major mergers (Naab et al. 2006), sequences of repeated minor mergers (Bournaud et al. 2007a), or merging galaxies embedded in a larger-scale group or cluster environment, showing a significant influence of the ram-pressure and large-scale gravity field on interacting galaxies (Kapferer et al. 2008, Martig \& Bournaud 2008).

We present here recent progress in terms of numerical resolution, in simulations of ``standard'' mergers of two massive spiral galaxies merging together, with no multiple mergers and no large-scale environment accounted for (see Section~2). While the remnants of major mergers in numerical simulations do globally resemble elliptical galaxies, whether the major merger scenario can really explain the exact properties of most real elliptical galaxies or not is still an open question (see Burkert et al. 2007). This raises the question of whether or not most simulations of galaxy mergers have a high-enough resolution to predict accurately the detailed properties of merger remnants: this is studied in Section~3, where we show that a too modest resolution may significantly affect the properties of numerical merger remnants. Our recent very-high resolution simulations also enable to study the smallest structures formed in galaxy mergers, like decoupled cores, tidal dwarfs, and super star clusters: results on these aspects will be presented in Section~4.

\section{The simulations and resolution}
 
 The simulations presented here were performed with a particle-mesh code based on a FFT Poisson solver for the gravity. The gas dynamics is modeled with a sticky-particle scheme, and a Schmidt law is used for star formation. A thorough description of the code and initial conditions can be found in Bournaud et al. (2008) and references therein. 
 The stellar disks have initial masses of $2 \times 10^{11}$~M$_{\odot}$ and we use Burkert (core) profiles for the dark halos. In the following we present simulations obtained at three different resolutions, all with the same 0.5~Gyr timestep:
\begin{itemize}
\item[--] the {\it low}  resolution has a softening of 180~pc (up to 25~kpc from each galaxy center, and degraded at larger distances) and uses $10^5$ particles/galaxy/component (the components are: stars, gas, and dark matter; this implies a total of $6 \times 10^5$ particles).
\item[--] the {\it medium}  resolution has a softening of 80~pc and $10^6$ particles/galaxy/component (total of $6 \times 10^6$ particles).
\item[--] the {\it high}  resolution has a softening of 32~pc and uses $6 \times 10^6$ particles/galaxy/component (total of $36 \times 10^6$ particles). Too our knowledge, this is the highest resolution ever achieved so far in this kind of simulations (see comparison with other works in Bournaud et al. 2008).
\end{itemize}

\section{Global properties of elliptical galaxies: a resolution study}

To test the standard proposal that ellipticals have been formed by galaxy mergers, and explore in more detail which sort of merger has formed them (binary major mergers, repeated minor or major merger, wet/dry mergers, etc..), recent works have measured in detail the properties of such numerical merger remnants (references above). This raises the question of whether or not numerical simulations with typically a few $10^5$ particles per galaxy have converged in terms of the detailed properties of these ``ellipticals''. 

Here we present a wet 1:1 merger model: the two spiral progenitors contain 83\% of stars and 17\% of gas (plus a dark halo with a Burkert profile), and they have the same mass. The simulation was run at the {\it low}, {\it medium}  and {\it high} resolutions defined above. The morphology of the relaxed merger remnant under the very same projection is shown for each resolution on Figure~1. Among the many quantitative parameters measured (which will all be presented in Bois et al. 2009, in preparation), we show on Figure~2 the radial variations of the robust tracer of angular momentum $\lambda_R$, defined by Emsellem et al. (2007), measured for a set of isotropically distributed projections. 

\begin{figure}  
\centering
\includegraphics[width=8cm]{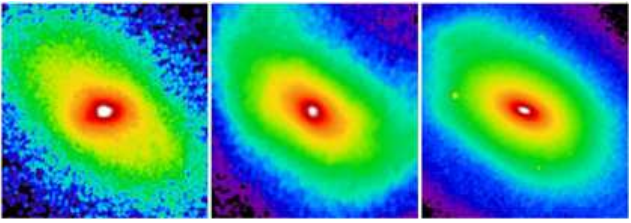}  
\caption{Projected stellar density maps for a major merger between two spiral galaxies. The {\it low}, {\it medium}  and {\it high}  resolution models of the same merger are viewed with the same projection angle. These snapshots show an area of 16x16~kpc.}  
\label{label_for_figure}  
\end{figure}

\begin{figure*}  
\centering
\includegraphics[height=3.4cm]{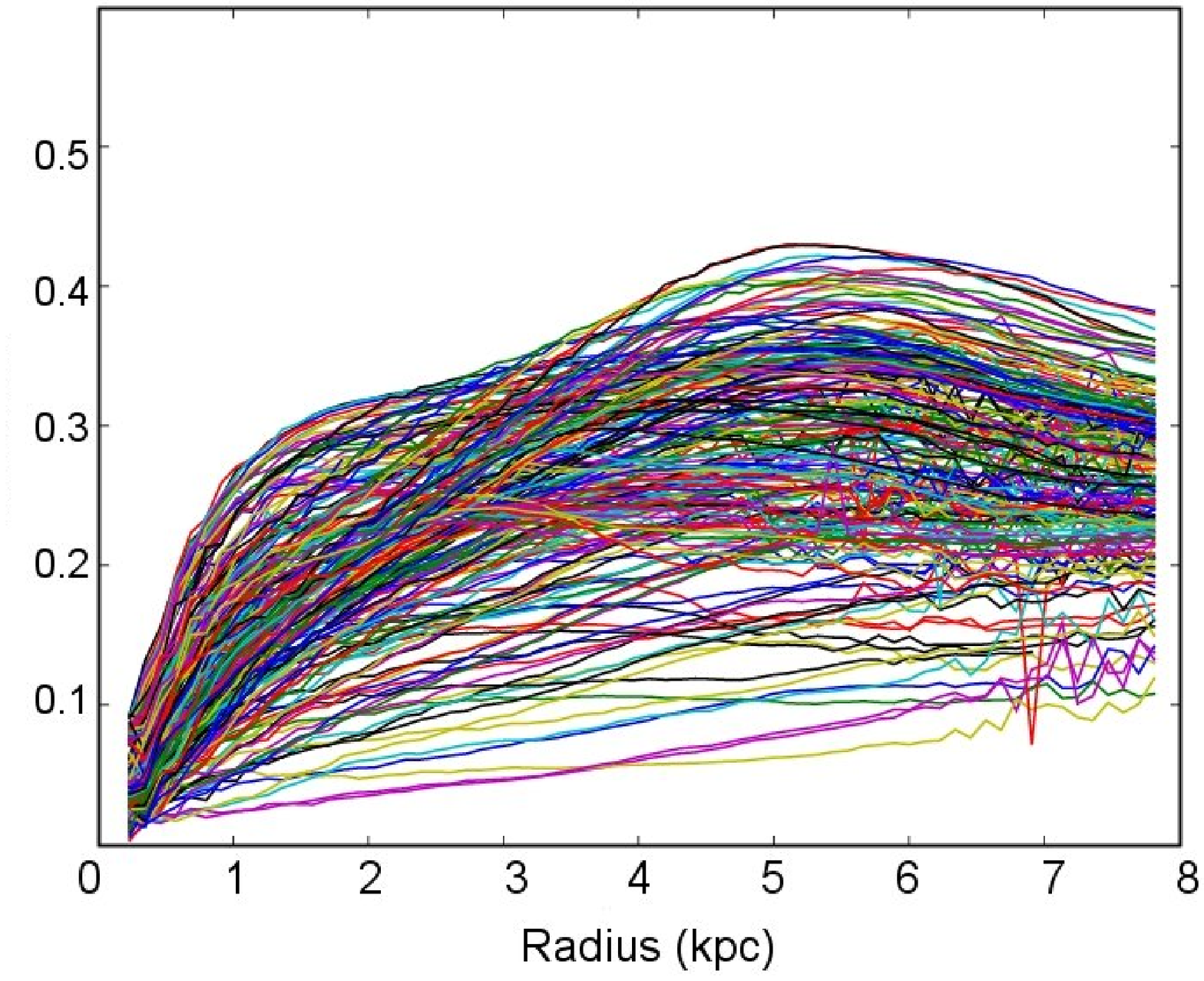}\hspace{.3cm}  
\includegraphics[height=3.4cm]{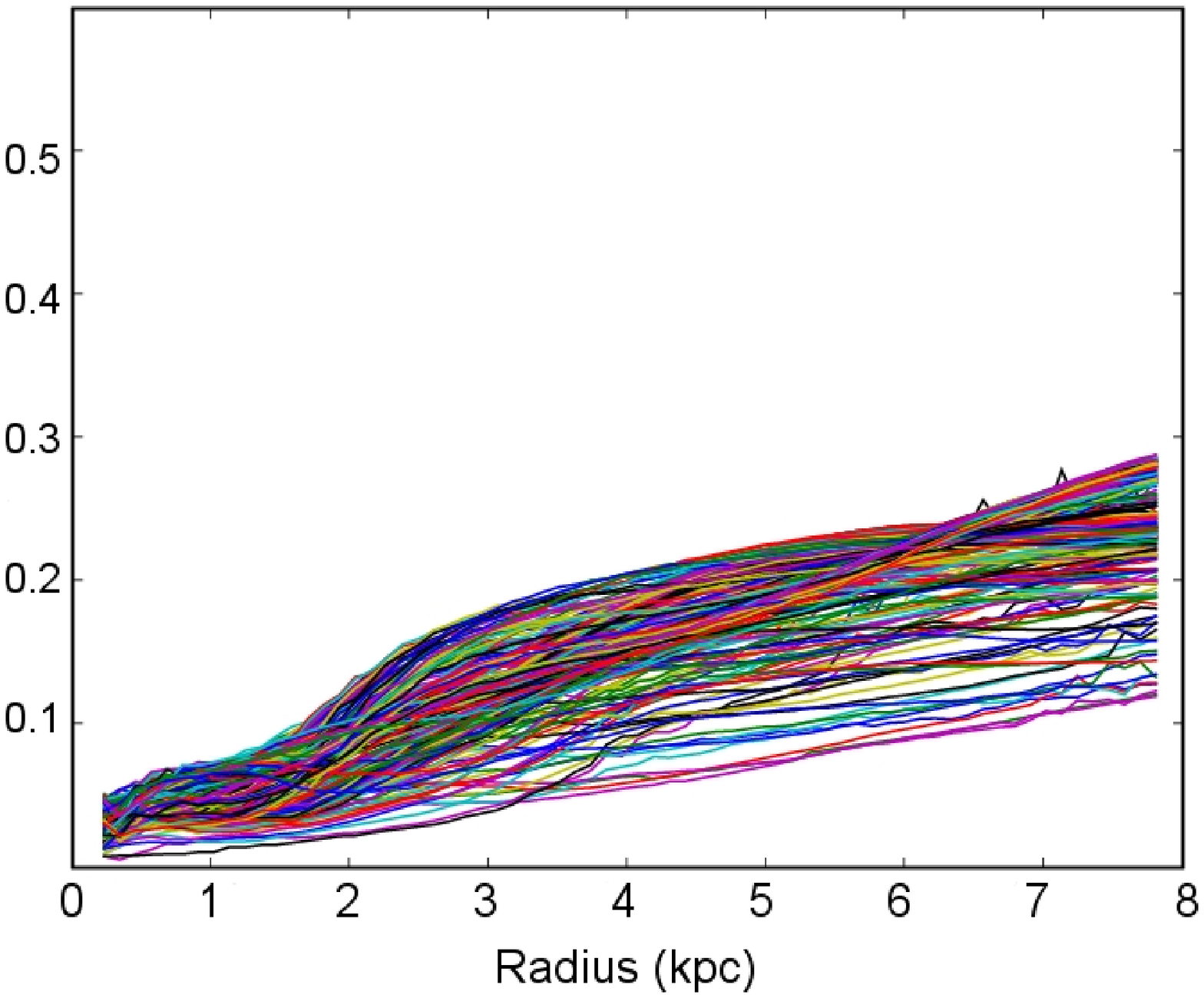}\hspace{.3cm}  
\includegraphics[height=3.4cm]{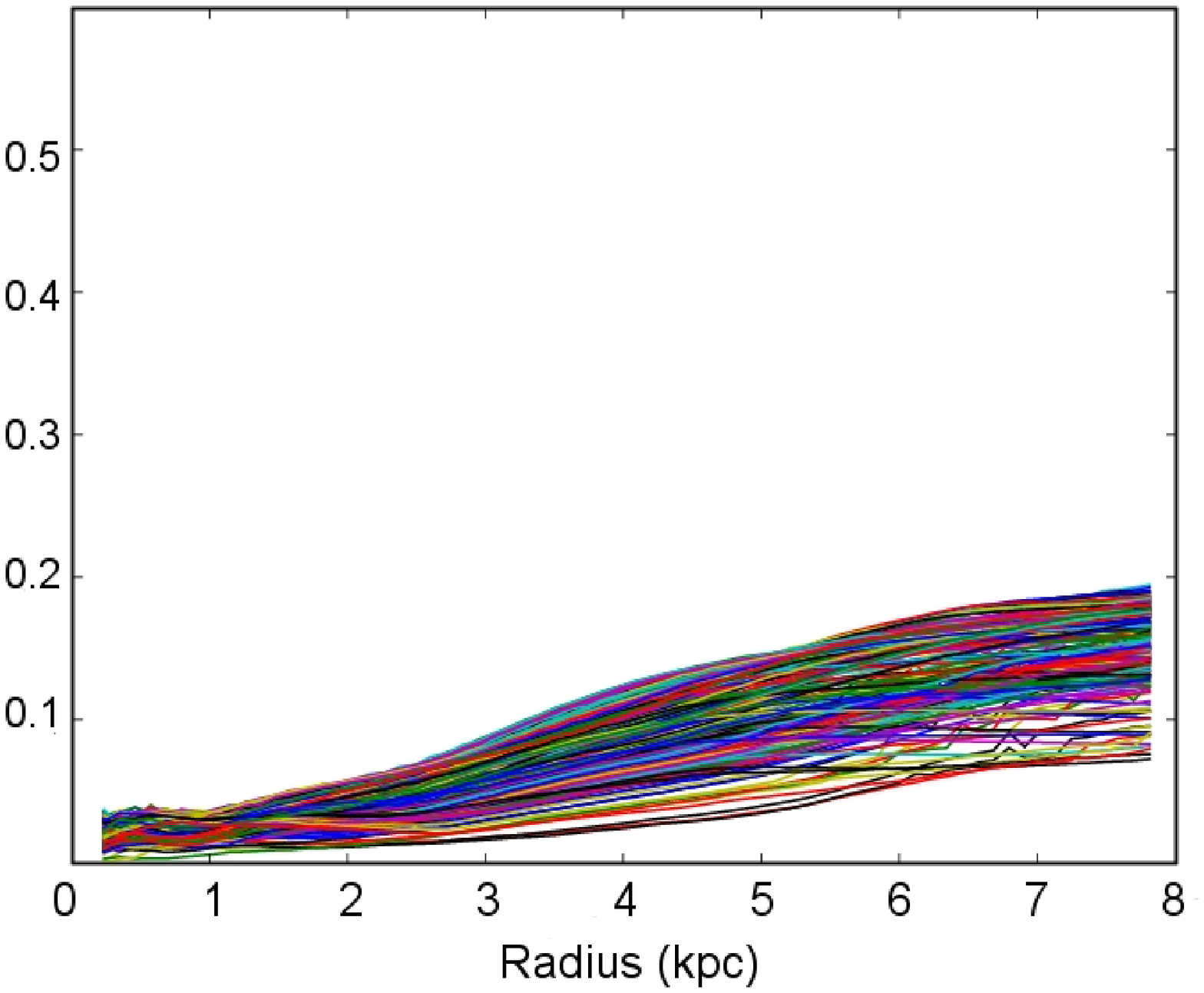}  
\caption{Radial profiles of the angular momentum tracer $\lambda_R$ of a relaxed merger remnant, for the {\it low}, {\it medium}  and {\it high}  resolution models respectively. The many curves in each plot correspond to different projection angle, that were chosen to be isotropically distributed.}  
\label{label_for_figure}  
\end{figure*}  

Our main conclusions follow, and we made sure (Bois et al. 2009) that these effects are not artifacts caused by somewhat different initial conditions (defining the very same initial conditions at different resolution is not trivial), do not affect only the gas dynamics and star formation but also the stellar and dark matter dynamics (sizeable results are found for dry mergers), and do not result simply from a change in the orbit (because the dynamical friction would not be equally resolved in the three cases), but really result from the way the violent relaxation processes are resolved:

\begin{itemize}
\item[--] At the three resolutions, the merger remnant resemble an elliptical galaxy, with about the same flattening (axis ratio) except in the central 3~kpc, and a similar $r^{1/4}$-like density profile. The effective (half-mass) radius is 2--2.5~kpc, depending on the projection angle, in the three cases.
\item[--] In more detail, significant variations can be seen in the isophotal shape (Fig.~1). The {\it low}  resolution case is overall disky with a major ($\sim70^{\mathrm{o}}$) isophotal twist. The {\it medium}  and {\it high}  resolution cases are boxy under the same projection, with little isophotal twist.
\item[--] Important kinematical variations are seen (Fig.~2), the angular momentum profile of the {\it low} resolution case differs dramatically from the {\it medium}  and {\it high}  resolution ones -- even the two later cases differ from each other, but in a more modest way and the shape of their $\lambda_R$ profiles is more comparable. In particular, within their effective radius, the {\it low}  resolution model would be classified as a fast-rotator (as defined by Emsellem et al. 2007) while the {\it medium}  and {\it high}  ones appear to be slow-rotators: this is a major difference as this typically defines two different classes of early-type galaxies (Emsellem et al. 2007), and the origin of the slowly-rotating ellipticals is a particular challenge for merger scenarios (as pointed out by Naab \& Ostriker 2007).
\end{itemize}

Overall, our {\it low}  resolution model (180~pc, $10^5$ particles/galaxy /component), which is not so ``low'' compared to recent works in the field, gives only a gross prediction of the properties of the merger remnant, and the merger/relaxation processes are insufficiently resolved to accurately predict the detailed properties of this ``elliptical''. Our {\it medium}  resolution (80~pc, $10^6$ part/gal/comp), which is quite high compared to several recent works, gives more accurate predictions of the morphology and kinematics. It generally compares favorably to the {\it high}  resolution case (32~pc, $6\times 10^6$ part/gal/comp -- the highest to date), but some results may still vary with resolution, in particular the advanced parameters used to deeply explore the structure of ellipticals like $\lambda_R$. This shows that resolution issues are a significant concern in attempts to compare in detail the properties of observed ellipticals with prediction of merger models, even in recent simulations with a rather high number of both stellar and gaseous particles.

\section{Structure formation in galaxy mergers: from decoupled cores to tidal dwarfs and globular clusters}

While the total stellar mass density of merger remnants is dominated by smooth spheroids (Fig.~1), interstellar gas fuels the formation of many substructures of gas and young stars. Kinematically decoupled cores (KDCs) are frequently seen in our medium- and high-resolution simulations of galaxy mergers (see Fig.~3). They form when a merger-driven gas inflow triggers the formation of a dense concentration of young stars inside, typically, the central kpc. Such stellar systems formed during/after the merger largely contribute to the total stellar kinematics in the central regions, while the larger-scale system is dominated by stellar populations older than the merger. Torques that act on the gas structures during the galaxy collision frequently cause the central component to have a kinematic axis misaligned with the older stellar system, and the KDC can even sometimes be counter-rotating w.r.t. the rest of the elliptical. Although our statistics are limited for the moment, we do not find KDCs to be preferentially associated to morphological decoupling like large isophotal twists. Note that major wet mergers are not necessarily the only mechanism producing kinematically decoupled components in early-type galaxies (Naab et al. 2007, di~Matteo et al. 2008).

\begin{figure}  
\centering
\includegraphics[width=3.5cm]{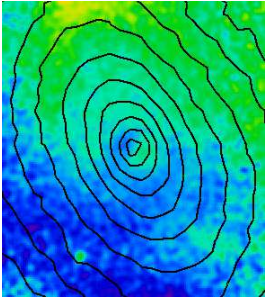}  
\caption{Stellar velocity field in a high-resolution wet 1:1 merger remnant, showing a counter-rotating KCD. The black contours show the projected stellar mass density (log spacing), this snapshot shows a 10x12~kpc area.}  
\label{label_for_figure}  
\end{figure}

\begin{figure}  
\centering
\includegraphics[width=7.5cm]{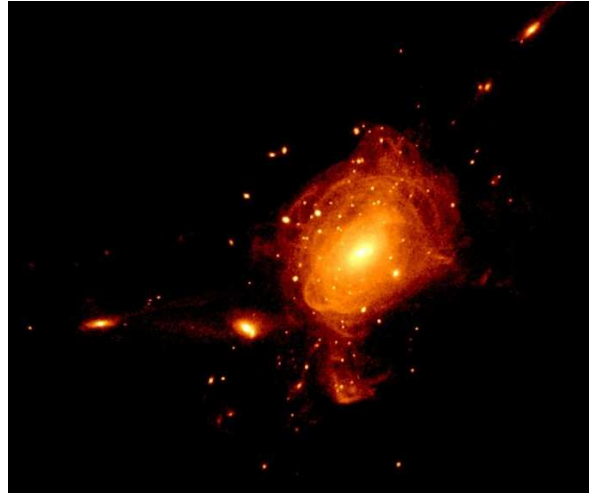}  
\caption{Projected mass density of the ``young'' stars, i.e. those formed during or after the merger, in a wet 1:1 high-resolution merger remnant.}  
\label{label_for_figure}  
\end{figure}

\begin{figure}  
\centering
\includegraphics[width=7cm]{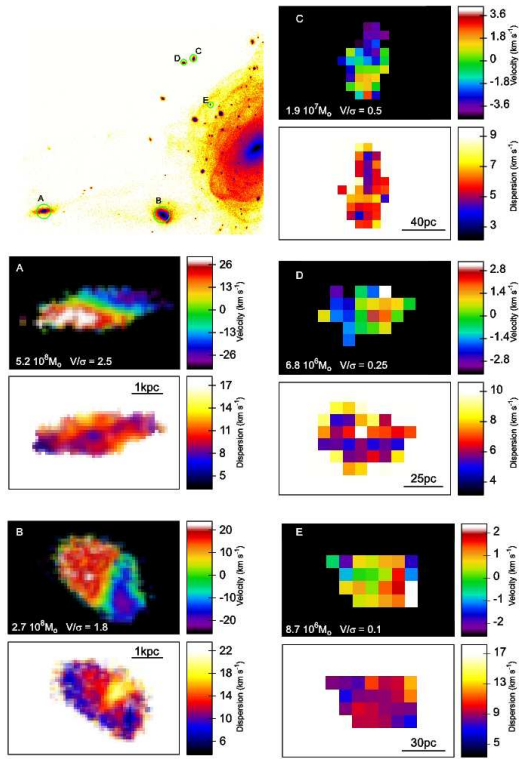}  
\caption{Line-of-sight velocity and dispersion fields for several young stellar objects formed in the mergers: two TDGs (A, B) and three SSCs/GCs (C, D, E).}  
\label{label_for_figure}  
\end{figure}  

The outskirts of merger remnants also contain a variety of young stellar substructures. Figure~4 displays the density map of the ``young'' stars formed during/after the merger, in a relaxed high-resolution merger remnant (see also Bournaud et al. 2008) : one can notice streams and shells, and more than one hundred of small stellar objects. The three most massive stellar objects have masses of $10^{8-9}$~M$_{\odot}$ radii of about a kpc and are supported by rotation ($V > \sigma$ -- see objects A and B on Fig.~5): they resemble Tidal Dwarf Galaxies (TDGs). The many other, smaller objects have lower masses, ranging from $10^{5}$~M$_{\odot}$ (the resolution limit) to a few $10^{7}$~M$_{\odot}$. They are very compact with radii from $\sim$30~pc (the softening limit) to $\sim$150~pc, and are tightly bound ($E_G/2E_K < -1$, see Bournaud et al. 2008), supported by random motions with little rotation ($V / \sigma < 1$, see Fig.~5). These Super Star Clusters (SSCs) are thus likely progenitors of Globular Clusters (GCs). The formation of such SSCs/GCs in major mergers had been predicted for instance by Li et al. (2004) and is here directly resolved.

\begin{figure}  
\centering
\includegraphics[width=6.5cm]{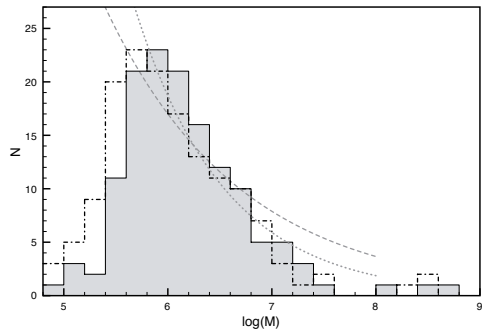}  
\caption{Mass function of the young bound stellar objects (SSCs/GCs and TDGs) formed in the major merger displayed on Fig.~4. A significant dip separates SSCs/GCs and the three massive TDGs. The dashed histogram corresponds to a test simulation with modified gas dissipation parameters (see Bournaud et al. 2008).}  
\label{label_for_figure}  
\end{figure}

TDGs are SSCs/GCs appear to be two different kind of objects: TDGs are large, rotating, found in the outer tidal tails. Their formation occurs when a giant molecular cloud or a dense region of the progenitor spiral disk is moved into a tidal tail by the interaction, following the mechanism proposed by Elmegreen et al. (1993). With large dark haloes, large regions of the outer disk can even be displaced into the outer regions of a tidal tail without being disrupted, and can form a new self-gravitating objects here (Duc et al. 2004).

The SSCs/GCs are formed in the outer tidal tails and other dense streams of gas at smaller radii, they have a much lower specific angular momentum and are more compact. This, together with a dip in the mass function between SSCs/GCs and TDGs (Fig.~6), implies that these are two different categories of objects, and the three large TDGs in our simulation are not just the high-mass end of the SSC distribution. The SSCs are formed by the fragmentation of dense gaseous structures, following the mechanism studied in particular by Wetzstein et al. (2007). In that vein, the ``tidal dwarfs'' in Wetzstein et al. models do not resemble the three massive TDGs in our model, but rather to the high-mass end of the more compact SSCs. This does not mean that they should never be considered as tidal dwarfs: indeed, with masses of a few $10^7$~M$_{\odot}$, they are really as massive as some dwarf galaxies. There could thus be two types of tidal dwarf galaxies: the high-mass end of SSCs/GCs, which would be compact spheroidal galaxies, and the larger, more massive and rotating TDGs, formed by a different mechanism and resembling the kpc-sized rotating TDGs observed in several interacting systems like NGC~5291 (Bournaud et al. 2007b). TDGs of this later type may still evolve into smaller dwarf spheroidals over a few Gyrs (Metz \& Kroupa 2007).

\section{Conclusions}

High spatial and mass resolution enables the formation of small objects to be resolved in numerical models of galaxy mergers. In recent simulations, even the internal properties of small objects can be resolved -- like the rotation curves of tidal dwarfs (see Bournaud et al. 2007b). The results of such high-resolution models is overall in agreement with theoretical expectations, in particular regarding the formation of a population of GCs during major wet mergers.

High resolution is also a must to study accurately the large-scale structure of spheroidal galaxies formed in major mergers. Our recent resolution study shows that even simulations with a total of one million particles and a spatial resolution of a hundred of parsec can still be significantly affected by the resolution. Whether major mergers can explain the detailed properties of elliptical galaxies (boxyness, angular momentum distribution, anisotropy, etc..) is still largely unsolved, and understanding which type of merger would have formed the various kind of ellipticals remains an important challenge.



\begin{thebibliography}{}
  \bibitem{}  Barnes, J.~E.\ 1992, ApJ, 393, 484 
  \bibitem{}  Bois, M. et al. 2009 in preparation
  \bibitem{}  Bournaud, F., Duc, P.-A., \& Emsellem, E.\ 2008, MNRAS, 389, L8 
  \bibitem{}  Bournaud, F., Jog, C.~J., \& Combes, F.\ 2005, A\&A, 437, 69 
  \bibitem{}  Bournaud, F., Jog, C.~J., \& Combes, F.\ 2007a, A\&A, 476, 1179 
  \bibitem{}  Bournaud, F., et al.\ 2007b, Science, 316, 1166 
  \bibitem{}  Burkert, A., Naab, T., \& Johansson, P.~H.\ 2007, ApJ in prep., arXiv:0710.0663 
  \bibitem{}  Cox, T.~J., Jonsson, P., Primack, J.~R., \& Somerville, R.~S.\ 2006, MNRAS, 373, 1013 
  \bibitem{}  di Matteo, P., Combes, F., Melchior, A.-L., \& Semelin, B.\ 2007, A\&A, 468, 61 
  \bibitem{}  di Matteo, P., Combes, F., Melchior, A.-L., \& Semelin, B.\ 2008, A\&A, 477, 437 
  \bibitem{}  Duc, P.-A., Bournaud, F., \& Masset, F.\ 2004, A\&A, 427, 803 
  \bibitem{}  Elmegreen, B.~G., Kaufman, M., \& Thomasson, M.\ 1993, ApJ, 412, 90 
  \bibitem{}  Emsellem, E., et al.\ 2007, MNRAS, 379, 401 
  \bibitem{}  Kapferer, W., Kronberger, T., Ferrari, C., Riser, T., \& Schindler, S.\ 2008, MNRAS, 923 
  \bibitem{}  Li, Y., Mac-Low, M.-M., Klessen, R. S., 2004, ApJL, 614, L29
  \bibitem{}  Martig, M., \& Bournaud, F.\ 2008, MNRAS, 385, L38 
  \bibitem{}  Metz, M., \& Kroupa, P.\ 2007, \mnras, 376, 387 
  \bibitem{}  Mihos, J.~C., \& Hernquist, L.\ 1994, ApJ, 437, 611 
  \bibitem{}  Naab, T., \& Burkert, A.\ 2003, ApJ, 597, 893 
  \bibitem{}  Naab, T., Khochfar, S., \& Burkert, A.\ 2006, ApJL, 636, L81 
  \bibitem{}  Naab, T., \& Ostriker, J.~P.\ 2007, arXiv:astro-ph/0702535 
  \bibitem{}  Naab, T., Johansson, P.~H., Ostriker, J.~P., \& Efstathiou, G.\ 2007, ApJ, 658, 710 
  \bibitem{}  Walker, I.~R., Mihos, J.~C., \& Hernquist, L.\ 1996, ApJ, 460, 121 
  \bibitem{}  Wetzstein, M., Naab, T., \& Burkert, A.\ 2007, MNRAS, 375, 805 
\end{thebibliography}
\end{document}